\documentstyle[12pt]{article}
\begin{document}
\baselineskip=18pt plus 2pt minus 2pt

\def\Refs{\list{}{\topsep-20pt\itemsep0pt\parsep0pt}\item[]}
\let\endRefs=\endlist
\def\References{\begin{center}
{\sc References}
\end{center}
\vspace*{-17.5pt}
\footnotesize}
\vspace{2mm}
\vspace{1.em}
\begin{center}
{\Large\bf LIENARD-WIECHERT POTENTIALS FOR MODEL OF EXTENDED SPACE}
\end{center}
\vspace{2mm}

\centerline{ \large D.Yu.Tsipenyuk and V.A.Andreev}
\vspace {2ex}
\vspace{2mm}
\vspace{2mm}
General Physics Institute of the Russian Academy of Sciences,
119991, Russia, Moscow, Vavilova str.38.
\vspace{2mm}

E-mail:  TSIP@KAPELLA.GPI.RU
\vspace{2mm}

\begin{center}
\begin{minipage}{135mm}
{ In the extended (1 + 4) -dimensional space $ G (T; \vec X, S) $ it is considered a model joining electromagnetic and gravitational fields. For the equations circumscribing these fields, the exact solutions adequate dot moving sources are found. Such solutions express through potentials Lienard-Wiechert in space $ G (T; \vec X, S) $. Their properties are analyzed.}

\end{minipage}
\end{center}
\vspace{1.em}
\vspace{1.em}

In works [1,2] was offered the model which extended special theory of relativity (STR) on 5-dimentional extended space
$G(T;\vec X,S)$ with metric
$$
(\;+,\;-,\;-,\;-,\;-\;).
$$

Within the framework of this extended space model (ESM) was considered possibility to join the gravitation and electromagnetism in one uniform field. The various details concern to the structure of the ESM is contained in works [3-7]. The experiment on check of predictions ESM [12,13,14] was made. For it quantitative analysis it is necessary to know exact solutions of the equations of the field, which were formulated in [1,2]. In the given work we shall found solutions, which appropriate to the point sources.

The principal difference ESM from STR is, that ESM allows to study processes during which the mass of a particle $m$ varies.
Under a mass of a particle $m$ we, following the recommendations of the review [8], shall understand it a rest-mass, which is a Lorentz scalar. In our model mass of a particle - component of a 5-vector in space $ G (T; \vec X, S) $, other components of this 5-vector are energy and impulse. Thus, the mass of a particle, generally speaking, varies during transformations in space $G (T; \vec X, S)$. But the charge, we assume, remains constant. We shall emanate also that the charge $ e $ is always connected to some mass $ m $ and we compare to a charged particle a parameter $ me $. As a charge $ e $ - scalar in extended space $ G (T; \vec X, S) $, parameter $ me $ - component of a vector in this space.
As the 5-th additional coordinate we propose to use the interval $ S $ which already exists in the Minkowski space.
$$
s^2\;=\;(ct)^2\;-\;x^2\;-\;y^2\;-\;z^2\;.
                                                           \eqno(1)
$$
This magnitude is saved in case of usual Lorentz transformations in Minkowski space $ M (T; \vec X) $, but varies in case of turns in extended space $ G (T; \vec X, S) $. Thus, Minkowski space $ M (T; \vec X) $ is a cone in extended space $ G (T; \vec X, S) $. A 5-vector in space $ G (T; \vec X, S) $ we shall designate by a hyphen $ (\bar r) $, vector in 4-d Euclidean space with coordinates $ (x, y, z, s) $.  Vector in it 3-d subspaces are designated by upper narrow and figure below, indicating dimensionality of space, to which this vector concerns: $ (\vec v _ {(3)}, (\vec v _ {(4)}) $. As a rule, from a context it is clear, about what of 3-d  subspace there is a speech and it data in the nomenclature of a vector we not indicate.
In work [1] was shown, that the uniform field joining a gravitation and an electromagnetism, is generated by a current, which in case of one particle is an isotropic 5-vector
$$
{\bar J}\;=\; (j_t,\vec j, j_s)\;=\;
\left(\frac{q c}{\sqrt{1-\beta^2}}\;, \frac{q {\vec u}}
{\sqrt{1-\beta^2}}\;,\;q c\;\right),
                                                             \eqno(2)
$$
where

$$
q\;=\;me,\;\;
q{\vec u}\;=\;me\vec u(t,x,y,z).
                                                                \eqno(3)
$$

If there are some particles, their common current is a sum of currents of separate particles. The additional transformations, which are available in extended space $ G (T; \vec X, S) $, change magnitude $ q $ similarly to that, as they change a rest-mass $ m $ [1,2,3].

In [1] the extended set of Maxwell equations circumscribing this field was constructed. As well as in case of a usual electrodynamics, the strengths express through the components of 5-vector potential
$$
\left(\varphi,{\vec A}, A_s\right)\;=\;
\left( A_t, A_x, A_y, A_z, A_s\right)\;.
                                                                \eqno(4)
$$
Potential (4) and current (2) are connected by the equations
$$
\Diamond  A_t\;=\;-\frac{4\pi}c j_t,                        \eqno(5)
$$
$$
\Diamond {\vec A}\;=\;-\frac{4\pi}c {\vec j},       \eqno(6)
$$
$$
\Diamond  A_s\;=\;-\frac{4\pi}c  j_s.              \eqno(7)
$$

Here
$$
\Diamond\;=\;\frac{\partial^2}{\partial s^2}+\frac{\partial^2}{\partial x^2}+
\frac{\partial^2}{\partial y^2}+
\frac{\partial^2}{\partial z^2}-\frac1{c^2}\frac{\partial^2}{\partial t^2}.
							        \eqno(8)
$$

To receive from system (5) - (7) usual set of equations for vector potential,
which is used in an electrodynamics, it is necessary to take into account, that
a mass $ m $ - scalar, concerning Lorentz transformations, to divide each of
the equations (5) and  (6) by this mass $ m $, and equation (7) divide by
charge $ e $, and also to take into account that now all magnitudes do not
depend on a variable $ s $. Therefore operator (8) turns in usual
dalembertian. After that the system (5) - (7) will come to a set of equations
for potentials of a usual classical electrodynamics and equation for a
gravitational field, which is a scalar on to Lorentz transformations. These
sets of equations look like $$ A_t\;=\;-4\pi\rho,
 \eqno(9) $$ $$ \vec A\;=\;-\frac{4\pi}c \vec \rho,                  \eqno(10)
$$
$$
  A_s\;=\;-\frac{4\pi}c \rho_s.                        \eqno(11)
$$
Here in right parts of equations (9) - (11) there are components of the current $$
\bar \rho\;=\;
\left(\frac{e c}{\sqrt{1-\beta^2}}\;, \frac{e\vec v}
{\sqrt{1-\beta^2}}\;,\;m c\;\right)\; \;.
                                                                 \eqno(12)
$$

The equations (9), (10) describe an electromagnetic field of a moving dot charge, and equation (11) scalar potentials of a gravitational field. Thus, in case, when the mass is fixed, i.e. there is no association from fifth coordinates $ s $, the gravitational and electromagnetic fields can be considered separately, dividing the equations (9), (10) and (11). If such association has a place, they should be considered as one uniform field.

For a set of equations of the Maxwell the major role is played by the solution called as Lienard-Wiechert potentials. This solution gives an obvious aspect of an electromagnetic field, which creates by a charge driven by any, but quite definitely way. The law of motion of a charge is the environmental condition, defining structure of a solution. In the given section we shall construct Lienard-Wiechert potentials for a system (5) - (7). On the formal structure they look like retarded potentials.
The idea of a construction of retarded potentials is based that the magnitude of strength of a field in a point $ (x, y, z) $ in an instant $ t $, created by a driven charge, is set by a position of a charge in some previous instant $ t ' < t $, which magnitude is determined by the law of motion of a particle.
In a usual electrodynamics the distance between a radiant of a field and view point is a distance in 3-d Euclidean space. In our model to two points in extended space $ G (T; \vec X, S) $ is compared 4-d space interval
$$
R_{(4)}^2\;=\;(x-x')^2+(y-y')^2+(z-z')^2+(s-s')^2,
							        \eqno (13)
$$
being a part of a full interval (1).

Let's discuss a problem - what parameters should be substituted in the formula (13) for calculation of the Lienard-Wiechert potentials.

Space coordinates $ (x, y, z), \; (x ', y ', z ') $ determine accordingly position of a point in which measure strength of a field in an instant $ t $ position of a charge in an instant $ t ' $.

In works [1,2,3] was shown, that fifth the coordinate $ s $ can be connected with an index of refraction $ n $. Distance by the 5-th coordinate $ (s-s') ^ 2 $ we shall define as follows. If the particle is move in blank space and this particle did not intersected fields with $ n> 1 $, then in (13) $ (s-s') = 0 $. If such sites meet, i.e. the segment connecting charge with a viewpoint lies in not empty 4-d Euclidean space with fields with $n>1$, the space interval describes by (13). The magnitude of a component $ (s-s') = R _ s $ depends not on a value of an index of refraction in these points, but from its value in all points along connecting this points path. Let's mark, that this path can and be not a rectilinear segment, it the geometry depends on distribution of factor of a refraction $ n (x, y, z) $. Let viewpoint of a field and source are from each other at 3-d distance
$$
R_{(3)}^2\;=\;(x-x')^2+(y-y')^2+(z-z')^2
							        \eqno(14)
$$
And light passes this distance during $ \Delta t $. Then under the interval definition
$$
(s-s')^2\;=\;c^2(\Delta t)^2\;-\;R_{(3)}^2\;=\;R_s^2.
							        \eqno(15)
$$
In this expression $ s $ and $ s' $ are not values of a parameter $ S $ in a view point of a field and in the source point. As the residual $ (s-s') $ characterizes all segment connecting radiant with a viewpoint, we shall use hereinafter for it label $ R _ s $. In these designations the velocity of a charge on a variable $ s $ is $ u _ s $, and it acceleration is $ \dot u _ s $.
The velocity $ v _ s $ of light distributions along a coordinate $ S $ is determined with the help of relation
$$
v_s^2\;=\;c^2-\frac{R_{(3)}^2}{(\Delta t)^2}.
							        \eqno(16)
$$

For the definition 4-d distance $ R _ {(4)} $ it is necessary magnitude (15) to substitute in the formula (13). 4-d distance $ R _ {(4)} $, speed of light $ c $, 3-d distance $ R _ {(3)} $ and speed of light in a medium $ v $ are connected one another by relation
$$
\frac{R_{(3)}}v\;=\;\frac{R_{(4)}}c
							        \eqno(17)
$$
or
$$R_{(3)}n = R_{(4)},$$
Where $ n $ - index of refraction of a medium.
Connection of a moment of an emission of a field $ t ' $ with a moment of it observation $ t $ is determined with the help of relations
$$
t'\;+\;\frac{R_{(4)}(t')}c\;=\;t,
							        \eqno(18)
$$
which is the same, as the similar relation in usual electrodynamics. The only difference is that instead of 3-d distance in this relation stands 4-d distance. The expression (18) is the equation on time of an emission $ t ' $, expressing it in time of observation $ t $. In a free space it has only one solution [9], however it is known, that in a medium it not and it can have few solutions [10]. We shall discuss this problem more in detail later.
Let's construct now potentials Lienard-Wiechert. For this purpose, as well as usually, at first we shall write potentials of a fixed charged particle, and then we shall rewrite them in the form, invariant concerning transformations in extended space $ G (T; \vec X, S) $.
If we have a dot particle with a charge $ e $ and mass $ m $. Appropriated to this particle 5-vector current
$$
\bar J\;=\;
em\bar\delta(\bar r-\bar r')\left(\frac{c}{\sqrt{1-\beta^2}}\;,
\frac{\vec u}{\sqrt{1-\beta^2}}\;,\;c\;\right)\; \;.
                                                              \eqno(19)
$$
Substituting these values of components of a current in right parts of the equations (5) - (7) we shall receive, that in the case of zero velocity of a particle ($ \vec u = 0 $) the potentials look like
$$
A_t\;=\;\varphi\;=\;\frac{em}{R_{(4)}(t')}, \;\;\;\;\vec A\;=\;0,\;\;\;\;
A_s\;=\;\frac{em}{R_{(4)}(t')}.
                                                              \eqno(20)
$$
Here $ R _ {(4)} (t ') $ - distance in 4-d space from a source of the field up to a view point. Using a relation (18) expressions for potentials can be written as
$$
A_t\;=\;\varphi\;=\;\frac{em}{c(t-t')}, \;\;\;\;\vec A\;=\;0,\;\;\;\;
A_s\;=\;\frac{em}{c(t-t')}.
                                                              \eqno(21)
$$
As the potential is a vector in extended space $ G (T; \vec X, S) $, expression (21) is easily generalized on case of moving particles.
$$
\bar A\;=\;\frac{em\bar u}{(\bar R,\bar u)}.
                                                              \eqno(22)
$$
Here $ \bar u (t ') $ - 5-vector of a velocity of a charged particle, $ \bar R (t ') $ a position vector connecting radiant of a field with a view point, and $ (\bar R, \bar u) $ - their scalar product. The values of both these vectors undertake in an instant $ t ' $.
$$
\bar R\;=\;\{c(t-t'),(\vec r-\vec r\,'(t')),R_s\},\;\;\;
                                                              \eqno(23)
$$
$$
\bar u\;=\;\{u_t(t'),u_x(t'),u_y(t'),u_z(t'),u_s(t')\}.
$$

Both these vector are isotropic. Structure of the radius vector $ \bar R $ is similar to vector of an energy-momentum-mass of a photon [1,2,3], and vector of a velocity $ \bar u $ similar to vector of an energy-momentum-mass of a massive particle. In blank space
$$
\bar R\;=\;\{c(t-t'),(\vec r-\vec r\,'(t')),0\},\;\;\;
\bar u\;=\;\left\{\frac{c}{\sqrt{1-\beta^2}},
\frac{\vec u}{\sqrt{1-\beta^2}},c\right\}.
                                                              \eqno(24)
$$

If we shell write by components scalar product $ (\bar R, \bar u) $ the expression (22) for Lienard-Wiechert potentials is
$$
A_t=\varphi(t)\;=\;\frac{emc}{cR_{(4)}(t')-
(\vec R_{(4)}(t'),\vec u_{(4)}(t'))}.
                                                              \eqno(25)
$$
Here values $ \vec R _ {(4)}, \vec u _ {(4)} $ designate space parts of 5-vectors
$ \bar R, \bar u $
$$
\vec R_{(4)}\;=\;((x-x'),(y-y'),(z-z'),(s-s')),\;\;\;
							        \eqno(26)
$$
$$
\vec u_{(4)}\;=\;(u_x, u_y,u_z,u_s)\;=\;(\dot x',\dot y',\dot z',\dot s'),
$$
And their scalar product $ (\vec R _ {(4)}, \vec u _ {(4)}) $ is
$$
(\vec R_{(4)},\vec u_{(4)})\;=\;(\vec R_{(4)},\vec{\dot R}_{(4)})\;=\;
							        \eqno(27)
$$
$$
-((x-x')\dot x'+(y-y')\dot y'+(z-z')\dot z'+(s-s')\dot s').
$$

We shall use definition
$$
{\cal L}_{(4)}\;=\;R_{(4)}(t')-
\frac1c(\vec R_{(4)}(t'),\vec u_{(4)}(t')).
                                                              \eqno(28)
$$
With the help of (28) the potentials (22) can be noted as
$$
A_t(t)\;=\;\frac{em}{{\cal L}_{(4)}(t')},
                                                              \eqno(29)
$$
$$
A_x(t)\;=\;\frac{emu_x}{c{\cal L}_{(4)}(t')},\;\;\;
A_y(t)\;=\;\frac{emu_y}{c{\cal L}_{(4)}(t')},
$$
$$
A_z(t)\;=\;\frac{emu_z}{c{\cal L}_{(4)}(t')},\;\;\;
A_s(t)\;=\;\frac{emu_s}{c{\cal L}_{(4)}(t')}.
$$

The parameters $ t $ and $ t ' $ are connected one another by relation (18).
Let's note that the found in such a way Lienard-Wiechert potentials in extended space $ G (T; \vec X, S) $ as a special case is given by usual Lienard-Wiechert potentials in a continuous medium [10,11]. Their connection is defined with the help of relation (17).
The potentials (29) are solutions of a system (5) - (7)
$$
f=f((\vec R_{4},\vec v_{4})).
$$
This not spherical-symmetrical solution, therefore for this statement is not spreaded the theorem that $ n $ -dimensional Laplacian has solution in form of
$$
f\;=\;r^{-n+2}, \;\;при \;n>2,\;и \;\;\; f\;=\;\ln r\;\;при \;n=2.
$$

That fact that for fixed particles the solution (29) gives usual Coulomb potential immediately follows from the formula (21). Really, solution (29) we have received by passing from a fixed particle to driven. Therefore, if we stopping particle we shall return from a common solution (29) to special solution (21). This solution can be rewrite in the form (20), which in case of blank space $ (v = c, \; R _ {3} = R _ {4}) $ gives usual Coulomb potential.
Let's found now strengths of fields appropriate to Lienard-Wiechert potentials (29). For this purpose we shall use the formula
$$
F_{ik}\;=\;\frac{\partial A_i}{\partial x_k}\;-
\;\frac{\partial A_k}{\partial x_i}\;,\;\;\;i,k=0,1,2,3,4
							        \eqno(30)
$$
which connecting strength with potentials.
At first we shall calculate strength of an electrical field $ \vec E $. For this purpose it is necessary to use the formula
$$
\vec E\;=\;-\frac1{mc}\frac{\partial \vec A}{\partial t}\;-
\;\frac1{m} grad A_t,
							        \eqno(31)
$$
similar on which calculates an electromagnetic field in a usual electrodynamics [9]. The difference is, that in the formula (31) right parts is divided by a rest-mass of a source of a current $ m $. In a usual electrodynamics such division is carried out for stages of a construction of a 4-vector of a current from a 4-vector of an energy - impulse, in our model we refer this procedure on a moment of an evaluation of strengths of an electromagnetic field. Let's mark, that so defined field $ \vec E $ is a 2-component tensor in space of the Minkowski $ M (T; \vec X) $, but it not a tensor in extended space $ G (T; \vec X, S) $. In this space tensors are the magnitudes (30), and the strength (31) will be transformed during rotations in space $ G (T; \vec X, S) $ under the much more complicated law.

Substituting here expressions for potentials (29), we shall receive
$$
\vec E\;=\;
\frac e{c^2{\cal L}_{(4)}^3}\left((c^2-\vec u_{(4)}^2)
\left(\vec R_{(3)} - \frac1c\vec u_{(3)}R_{(4)}\right) +
\left(\vec R_{(3)} - \frac{\vec u_{(3)}}cR_{(4)}\right)
(\vec R_{(4)},\vec{\dot u}_{(4)})\right) -
$$
$$
\frac e{c^2{\cal L}_{(4)}^3}\left(\left(R_{(4)}^2-\frac{R_{(4)}}c
(\vec R_{(4)},\vec{u}_{(4)})\right)\vec{\dot u}_{(3)}\right)       .
							        \eqno(32)
$$

The magnetic field has form
$$
\vec H \;=\;\frac1m rot\vec A.
							        \eqno (33)
$$

Substituting here values of potentials, we obtain
$$
\vec H\;=\;-\frac e{c^2{\cal L}_{(4)}^2}\left[\vec R_{(3)},\vec{\dot u}_{(3)}
\right]\;+\;
\frac e{c^2R_{(4)}{\cal L}_{(4)}^2}\times
$$
$$
\left[\vec u_{(3)},
\left(c\vec R_{(3)} -
R_{(4)}\vec u_{(3)} +\left(c(\vec R_{(4)},\vec{u}_{(4)}) +
R_{(4)}(\vec R_{(4)},\vec{\dot u}_{(4)})-R_{(4)} u_{(4)}^2\right)
\frac{\vec R_{(3)}}{c^2{\cal L}_{(4)}}\right)\right].
							        \eqno(34)
$$

As well as usually, the fields $ \vec E $ and $ \vec H $ are connected by a relation
$$
\vec H\;=\;\frac1{R_{(4)}}[\vec R_{(3)},\vec E]\;.
							        \eqno(35)
$$
The formulas (32) - (35) are similar to the formulas for electric and magnetic fields in case of a usual electrodynamics [9]. The difference is, that into the formulas (32) - (35) enter as 3-d vector $ \vec R _ {3} $, and the module 4-d vector $ R _ {4} $, in usual electrodynamics in the appropriate formulas instead of the module 4-d vector $ R _ {4} $  stands the module 3-d vector $ R _ {3} $.

The formulas (32), (34) give expressions for electrical and magnetic fields, but with their help it is possible to receive expressions and for all other strengths (30). For this purpose it is necessary to take into account, that vector $ \vec u _ {(3)}, \vec R _ {(3)} $ in the formulas (32), (34) is 3-d vector, which components are selected from the 4-th components 4-d vectors $ \vec u _ {(4)}, \vec R _ {(4)} $. Such triples of components can be selected by a different mode. If we have selected components $ x, y, z $, we shall receive electrical and magnetic fields, which differ from usual fields only by that now their components depend also on a variable $ s $.
If we shall select other triple of variables, we shall receive other components of a tensor of strength (30). So for example, if we have selected variables $ y, z, s $, in the formulas (31), (33) there will be a 3-vector potential $ \vec A _ {(3)} $ with components $ (A _ y, A _ z, A _ s) $. Accordingly and in the left part of the formula (31) there will be same components of a 4-vector $ \vec E _ {(4)} $: $ \; \; (E _ y, E _ z, E _ s = Q = F _ {40}) $. The formula (33) gives in this case strengths $ H _ x, G _ y = F _ {42}, G _ z = F _ {43} $. The new fields $ Q, \vec G $ occur because now there are 5 potentials $ (A _ t, \vec A, A _ s) $ and all of them depend also on a new variable $ s $.
Let's write these fields
$$
Q\;=\;
\frac e{c^2{\cal L}_{(4)}^3}\left((c^2-\vec u_{(4)}^2)
\left(R_s - \frac1c u_sR_{(4)}\right) +
\left(R_s - \frac{u_s}cR_{(4)}\right)
(\vec R_{(4)},\vec{\dot u}_{(4)})\right) -
$$
$$
\frac e{c^2{\cal L}_{(4)}^3}\left(\left(R_{(4)}^2-\frac{R_{(4)}}c
(\vec R_{(4)},\vec{u}_{(4)})\right){\dot u}_s\right)       .
							        \eqno(36)
$$

To find components of a vector $ \vec G $, we shall introduce follow designations
$$
\vec D\;=\;(H_x,G_y,G_z),\;\;\vec R_{(3)}^{(x)}\;=\;(R_y,R_z,R_s),\;\;
\vec u_{(3)}^{(x)}\;=\;(u_y,u_z,u_s).
							        \eqno(37)
$$
The  superscript $(x)$ of the 3-d vectors specifies that coordinate 4-d space,
which is absent in this 3-d vector.

Using these designations we shall receive expression for a vector $ \vec D $
$$
\vec D\;=\;-\frac e{c^2{\cal L}_{(4)}^2}\left[\vec R_{(3)}^{(x)},
\vec{\dot u}_{(3)}^{(x)}
\right]\;+\;
\frac e{c^2R_{(4)}{\cal L}_{(4)}^2}\times
$$
$$
\left[\vec u_{(3)}^{(x)},
\left(c\vec R_{(3)}^{(x)} -
R_{(4)}\vec u_{(3)}^{(x)} +\left(c(\vec R_{(4)},\vec{u}_{(4)}) +
R_{(4)}(\vec R_{(4)},\vec{\dot u}_{(4)})-R_{(4)} u_{(4)}^2\right)
\frac{\vec R_{(3)}^{(x)}}{c^2{\cal L}_{(4)}}\right)\right].
							        \eqno(38)
$$

If in the formula (38) to replace a superscript $(x)$ on $ (y) $, it instead of
vector $ \vec D $ with components (37) will give a vector $ \vec D ' $ with
components
$$ \vec D'\;=\;(H_y,G_z,G_x).\;\;\ \eqno(39) $$
Here we take into account that
$$ \vec R_{(3)}^{(y)}\;=\;(R_z,R_s,R_x),\;\; \vec
u_{(3)}^{(y)}\;=\;(u_z,u_s,u_x).  $$

Thus, we see, that in dependence from what components of the 4-d vectors $ \vec u _ {(4)}, \vec R _ {(4)} $ are included into the formulas (32), (34), they allow to calculate various gangs of components of strength tensor (30).
Let's consider now concrete example of a particle with a mass $ m $ and charge $ e $, with fixed space coordinates $ x, y, z $, but the coordinate $ s $ can vary. This particle has follows 4-d state, velocities and acceleration vectors
$$
\vec R_{(4)}=(x,y,z,R_s)\;,\;\;\;\vec u_{(4)}=(0,0,0,u_s)\;,\;\;\;
\vec {\dot u}_{(4)}=(0,0,0,\dot u_s)\;.
							        \eqno(40)
$$
Here we use the entered earlier designation $ R _ s $, for a distance in 4-d Euclidean space between source of the field and view point.
Let's mark, that for charge moving with a velocity $ u _ s $ the modification of the distance $ R _ s $ happens, but the connection between them is rather complicated and is determined by distribution of an index of refraction $ n (\vec r) $ in all space between them. For it evaluation should be taken into account relations (16) - (18) and property of a medium.
Using the formula (32), we obtain expression for electric field strengths
$$
\vec E\;=\;ec\frac{c^2-u_s^2+R_s\dot u_s}{(cR_{(4)}-R_s u_s)^3}\vec R_{(3)}\;.
							        \eqno(41)
$$

The denominator of this expression always is more than zero
$$
cR_{(4)}-R_s u_s\;>\;0,
							        \eqno(42)
$$
as the velocity of the charge $ u _ s $ always is less than speed of light in a hollow $ c $, and the distance $ R _ s $ along the coordinate $ s $ always is less than the full distance $ R_ {(4)} $, due to the source and view point always are divided spatially.
But the numerator of expression (41) can change a sign in an association with magnitude and sign of acceleration $ \dot u _ s $. If this acceleration is rather great, i.e. the velocity of a modification of an optical density of a medium around fixed charge is great, the voltage of an electrical field $ \vec E $ will interchange the sign. Thus the of the charges with the same signs will begin to be attracted, and charges with opposite signs to be repelled.
It is interesting to compare expression (41) with expression for electric field strength, which is received with the help of usual Lienard-Wiechert potential [12]. If the 3-d velocity is directed along an axes $ X $
$$
\vec u_{3} \;=\;(u_x,0,0).
							        \eqno(43)
$$
In this case electric field component $ E _ x $
$$
E_x\;=\;\frac e{c^2{\cal L}_{(3)}^3}\left((c^2-u_x^2)\left(R_x-\frac{u_x}c
R_{(3)}\right)\;-\;(R_y^2+R_z^2)\dot u_x\right).
							        \eqno(44)
$$
Here magnitude $ {\cal L} _ {(3)} $ are similar to (28), only instead 4-d vectors there are 3-d vectors
$$
{\cal L}_{(3)}\;=\;R_{(3)}(t')-
\frac1c(\vec R_{(3)}(t'),\vec u_{(3)}(t')).
$$

From the formula (44) it is visible, that the component of strength $ E _ x $ can change the sign in an association with magnitude and sign of acceleration $ \dot u _ x $. However, in practice such modification of a sign, i.e. changing of an attraction in repulsion and on the contrary is rather inconveniently to observe. This is due to the superlarge magnitudes acceleration can be created only on very short time intervals. And this acceleration should be positive, i.e. the velocity of the charged particle must sharply increase.
The effect of change of a sign of a force of interaction arises only in that case, when the velocity of a particle is directed not on a line connecting particle and view point. In other words, the magnitude $ (R _ y ^ 2 + R _ z ^ 2) $ in the formula (44) should not be equal to zero. From here follows, that in the case of frontal motion of charged particles to each other changes of a sign of a force of interaction does not happen. This case has place only, when the aim parameter is not equal to zero. Such situation can arise, for example, in the case of falling of a charged particle on the Coulomb center. Near to this center the strength of a field is so great that can cause large acceleration of the incident particle and change of an attraction by repulsion.
Let's return to the analysis of strength (30) under condition of (40).
The field $ Q $ has an view
$$
Q\;=\;\frac e{c^2{\cal L}_{(4)}^3}\left((c^2-u_s^2)\left(R_s-\frac{R_s}c
R_{(4)}\right)\;-\;(R_x^2+R_y^2+R_z^2)\dot u_s\right).
							        \eqno(45)
$$
From the formula (45) it is visible, that a field $ Q $ also, as well as the field $ \vec E $, can change the sign in an association with a sign and magnitude of acceleration $ \dot u _ s $.
As the strengths $ \vec H, \vec G $ express through fields $ \vec E, Q $ with the help of of formulas (35), (38) they also change the sign in that moment, when it is done by fields $ \vec E, Q $.
Such changing of a field strength sign and, as a result, change of a Lorentz force sign it is possible to connect with a response of a radiation of these fields, which arises, when charged particles moves with acceleration.
The results obtained in the given work, show, that in the case of generalization of an electrodynamics on extended space it is possible to save all basic elements of the theory. The fields of dot charges are described by generalized Lienard-Wiechert potentials, which allow analyzing their behavior in a series of concrete cases. This result defines connection between electromagnetic and gravitational fields and determines direction, in which it is necessary to move with the purpose of a construction of their share theory.

\newpage

\vspace {2ex}

\centerline{ \large\bf References}

\begin{Refs}

\item[{[1]}] D.Yu.Tsipenyuk and V.A.Andreev,  {\it Kratkie soobstcheniya po fizike}
No 6, p. 23-34 (2000)  ({\it Bulletin Lebedev Physics Institute (Russian
      Academy of Sciences)}, Allerton Press, Inc., N.Y,2001);
gr-qc/0106093

\item[{[2]}] D.Yu.Tsipenyuk and V.A.Andreev, {\it Issledovano v Rossii} (Russian
electronic journal),  60 (1999);

      http://zhurnal.mipt.rssi.ru/articles/1999/060.pdf

\item[{[3]}] D.Yu. Tsipenyuk , V.A.Andreev {\it Structure of extended space}  Preprint IOFAN (General Physics Institute), {\bf5}, 25p., (1999).

\item[{[4]}] D.Yu. Tsipenyuk , V.A.Andreev, {\it Electrodinamics in extended space}  Preprint IOFAN (General Physics Institute) , {\bf9}, 26p., (1999).

\item[{[5]}] D.Yu. Tsipenyuk , V.A.Andreev, {\it Interaction in extended space}  Preprint IOFAN (General Physics Institute), {\bf2}, 25p., (2000).

\item[{[6]}] D.Yu. Tsipenyuk , V.A.Andreev, {\it Lienard-Wiechert potencials and Lorentz force in extended space}  Preprint IOFAN (General Physics Institute), {\bf1}, 20p., (2001).

\item[{[7]}] D.Yu. Tsipenyuk , V.A.Andreev, {\it Gravitational effects in extended space}  Preprint IOFAN (General Physics Institute), {\bf4}, 20p., (2001).

\item[{[8]}] L.B.Okun, {\it Usp. Fiz. Nauk}, {\bf158}, 511 (1989).

\item[{[9]}] L.D.Landau, E.M.Lifshits, {\it The classical theory of
fields}, M.:  Nauka, (1967).

\item[{[10]}] B.M.Bolotovskii, V.P.Bykov, {\it Usp. Fiz. Nauk},
{\bf 160}, 511 (1990).

\item[{[11]}] V.L.Ginzburg, {\it Theoretical physics and
astrophysics}, М., Nauka, (1981).

\item[{[12]}] Tsipenyuk D.Yu, {\it Kratkie soobstcheniya po fizike}
No 7, pp. 39-49 (2001)  ({\it Bulletin Lebedev Physics Institute (Russian
Academy of Sciences)}, Allerton Press, Inc., N.Y,2002).

physics/0107007.

\item[{[13]}] D.Yu.Tsipenyuk, {\it Gravitation\&Cosmology} Vol 7,
No.4(28), pp. 336-338 (2001).

physics/0203017

\item[{[14]}] D.Yu.Tsipenyuk, {\it Proc. of the XVI Int. Workshop on High
Energy Physics and Quantum Field Theory}. Eds.: M.N.Dubinin, V.I.Savrin - M.:
UNCDO publishing, 2002, pp. 398-405.

http://theory.sinp.msu.ru/~qfthep/2001/Proc2001.html

\end{Refs}

  \end{document}